\documentclass[aps,prl,twocolumn,groupedaddress,letterpaper]{revtex4}
\usepackage{graphicx}
\usepackage{amsmath, amssymb}
\usepackage{color}
\usepackage{graphicx}
\usepackage{multirow}
\usepackage{booktabs}
\usepackage{todonotes}
\pdfoutput=1

\def\ket#1{{\lvert}#1\rangle}

\def\stair{\lrcorner \hspace{-1.25 pt} \ulcorner} 
\begin{document}
\title{Experimental Demonstration of Blind Quantum Computing} 
\author{Stefanie Barz$^{1,2}$, Elham Kashefi$^{3}$, Anne Broadbent$^{4,5}$, Joseph F. Fitzsimons$^{6,7}$, Anton Zeilinger$^{1,2}$, Philip Walther$^{1,2}$}
 \affiliation{
 $^1$~Faculty of Physics, University of Vienna, Boltzmanngasse 5, A-1090 Vienna, Austria\\
 $^2$~Institute for Quantum Optics and Quantum Information (IQOQI), Austrian Academy of Sciences, Boltzmanngasse 3, A-1090 Vienna, Austria\\
 $^3$ School of Informatics, University of Edinburgh, 10 Crichton Street, Edinburgh EH8 9AB, UK\\
 $^4$ Institute for Quantum Computing, University of Waterloo,   200~University Avenue West, Waterloo, Ontario, Canada \mbox{N2L 3G1} \\
 $^5$ Department of Combinatorics \& Optimization, University of Waterloo, 200 University Avenue West, Waterloo, Ontario, Canada \mbox{N2L 3G1}\\
 $^6$ Centre for Quantum Technologies, National University of Singapore, Block S15, 3~Science Drive~2, Singapore~117543\
 $^7$ School of Physics, University College Dublin, Belfield, Dublin 4, Ireland}

\begin{abstract}
Quantum computers, besides offering substantial computational speedups, are also expected to provide the possibility of preserving the privacy of a computation. 
Here we show the first such experimental demonstration of blind quantum computation where the input, computation, and output all remain unknown to the computer. 
We exploit the conceptual framework of measurement-based quantum computation that enables a client to delegate a computation to a quantum server. 
We demonstrate various blind delegated computations, including one- and two-qubit gates and the Deutsch and Grover algorithms. Remarkably, the client only needs to be able to prepare and transmit individual photonic qubits.
Our demonstration is crucial for future unconditionally secure quantum cloud computing and might become a key ingredient for real-life applications, especially when considering the challenges of making powerful quantum computers widely available.
\end{abstract}

\maketitle
Among many quantum-enhanced applications, quantum computing has
generated much interest due to the discovery of
applications~\cite{Feynman1982, Deutsch1992, Grover1996, Shor1997} that
outperform their best-known classical counterparts. Although vast
technological developments already allow for small-scale quantum
computers with ionic~\cite{Cirac1995, Monroe1995, Kielpinski2002, Kim2010}, photonic~\cite{OBrien2003,Walther2005a,
Kiesel2005, Prevedel2007a,Lu2007a, Tokunaga2008, Kaltenbaek2010,
Vallone2010}, superconducting~\cite{Makhlin2001, Yamamoto2003,
Neeley2009, DiCarlo2009, Bialczak2010}, and solid
state~\cite{Berezovsky2008, Fushman2008, Hanson2008} systems, the
hurdles encountered in realizing quantum devices are enormous. This
intrinsic technical complexity may result in, initially, only a few
powerful quantum computers, or quantum servers, operating at
specialized facilities. Obviously, a key challenge in using such
central quantum computers is enabling a quantum computation on a
remote server, while keeping the client's data hidden from the
server~\cite{Childs2005, Arrighi2006,Giovannetti2008, DeMartini2009,
Broadbent2009, Aharonov2010}.

The classical analogue of this issue was addressed for the first time in 1978 by Rivest and co-authors~\cite{Rivest1978} and became 
one of the most
active fields in cryptography.
A full solution was over 30 years
in the making and enables~\cite{Gentry2009} the evaluation
of data-processing circuits over encrypted data without the need for
any decryption, but provides only computational security. In analogy to many widely used cryptographic protocols, this means that the
security relies 
on the assumption of a limit to the adversary's computational power, as well as on the difficulty of the underlying mathematical problem.

Remarkably, the recent theoretical work by Broadbent, Fitzsimons, and
Kashefi~\cite{Broadbent2009} overcomes this limitation and shows that
quantum computers can provide unconditional security in data
processing --- a hitherto unrecognized potential of quantum computers
that is not known to be achievable classically.
This new fundamental advantage of quantum computers is manifested in
the blind quantum computing (BQC) protocol that combines notions of
quantum cryptography and quantum computation to achieve the delegation
of a quantum computation from a client with no quantum computational
power to an untrusted quantum server, such that the client's data
remains perfectly private.

BQC uses the concept of  one-way quantum
computing~\cite{Raussendorf2001, Raussendorf2003, Danos2007,
Gross2007, Briegel2009}, a measurement-based model of
computation~\cite{Gottesman1999,Knill2001} which represents a paradigm
shift in the understanding of complex data processing by clearly
separating
the classical and quantum parts of a computation. In
the most general case, a one-way quantum computer is based on highly
entangled multi-particle states, so-called cluster
states, which are a resource for universal quantum computing. On these
cluster states, adaptive single-qubit
measurements alone are
sufficient to implement deterministic universal quantum computation.
Different algorithms require only a different pattern of single-qubit
measurements on a sufficiently large cluster state.

\begin{figure}
\includegraphics[width=0.95\columnwidth]{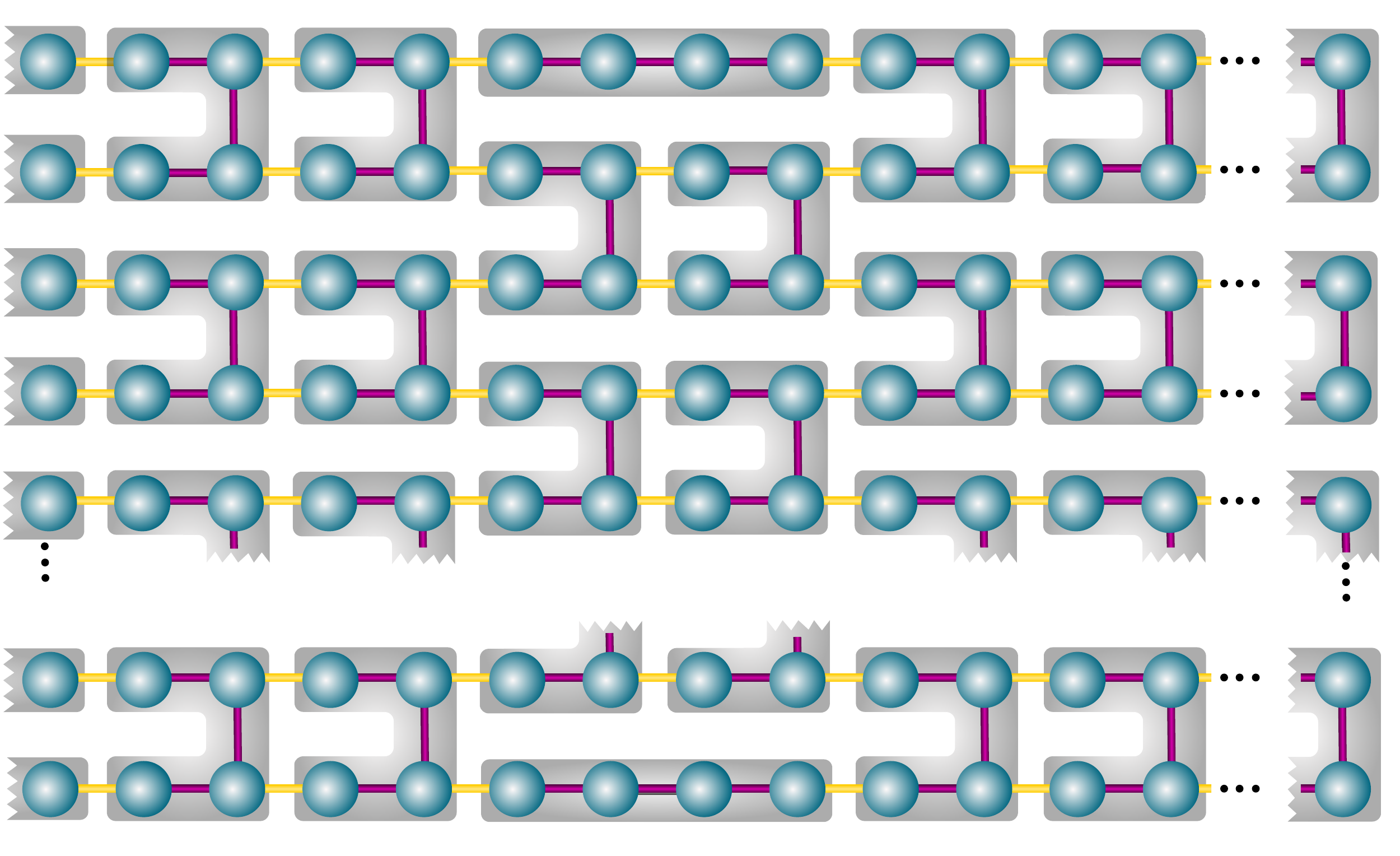}
\caption{\label{figure1}The universal blind cluster state for blind quantum computing. This family of cluster states can be built by joining (yellow edges), for example using optical fusion operations, smaller cluster states (purple edges, grey background)  that are in one of the configurations of $\ket{\Phi^{\hat{\theta}}}$ as implemented in the laboratory. The resulting state allows universal blind quantum computation when combined with measurements in the basis $\ket{\pm_\delta}, \delta \in \{0,\pi/4, \ldots 7\pi/4\}$.}
\end{figure}

Therefore, a
quantum computation is hidden as long as these measurements are successfully hidden. 
In order to achieve
this, the BQC protocol exploits special resources called \emph{blind
cluster states}
that must be chosen carefully to be a generic structure that reveals nothing about the underlying computation (see Figure~\ref{figure1}).
These blind cluster states are multi-particle entangled states created by
preparing qubits in  $\ket{\theta_j} = 1/\sqrt{2}(\ket{0} +
e^{i\theta_j}\ket{1})$, where $\ket{0}$ and $\ket{1}$ are the
computational basis of the physical qubits and $\theta_j$ is chosen
uniformly at random from $\{0, \pi/4, \ldots ,7\pi/4 \}$, and then interacting
each qubit via controlled-phase (CPhase) gates with its nearest
neighbours (here, $\text{CPhase} \ket{i}\ket{j}\mapsto
(-1)^{ij}\ket{i}\ket{j}$ with $i,j \in \{0,1\})$. 
Similar to the one-way quantum computer, a blind computation is described by a  \emph{pattern} of consecutive adaptive single-qubit measurements. Measuring the first qubit, initially in state~$\ket{\theta_1}$,
of a one-dimensional linear blind cluster in the basis $\ket{\pm_{{\delta_1}}}= 1/\sqrt{2}(\ket{0} \pm e^{i{\delta_1}}\ket{1})$ has the effect of applying a single-qubit rotation $R_z(-\delta_1 + \theta_1)$, on the encoded input state $\ket{+}$, followed by a Hadamard, $H$.
As long as the angle $\theta_1$ of the rotated qubit is unknown, the real rotation remains secret.
Here, $R_z(\phi) =  \exp(-i\phi\sigma_z/2)$,  $H=(\sigma_x + \sigma_z)/\sqrt{2}$ and $\sigma_x, \sigma_y$ and $\sigma_z$ denote the usual Pauli matrices.  
%

This feature of blind cluster states is used to perform a delegated computation on a server, such that all data and the whole computation remain hidden.
The only quantum power that is required from the client is the preparation of each qubit~$j$ in a state $\ket{\theta_j}$ and the transmission of the qubits to the server --- in particular, there is no need for any quantum memory~\cite{Hedges2010} or ability to perform quantum gates. From this point on in the protocol, the client communicates only  measurement instructions and can be considered completely classical. The quantum server, which \textit{can} perform universal quantum computation, performs a $\text{CPhase}$ gate between qubits received from the client.
Then in each round of interaction, the server performs adaptive single-qubit measurements in the $\ket{\pm_{\delta_j}}$ basis, as instructed by the client. The measurement basis is chosen such that ~$\delta_j=
\phi_j + \theta_j + \pi r_j$, where~$\phi_j$ is the desired target rotation and ~$r_j$ is a randomly chosen value
in~$\{0,1\}$ which hides the value of the measurement outcome.
These classical measurement angles are set in such a way to compensate for
the initial random rotation  $\theta_j$ and any other Pauli
byproducts~\cite{Danos2006, Prevedel2007a} produced by
previous measurements. 

In the present work, we present a optimised version of the original protocol using photonic qubits.
Photons are ideally suited for BQC as they provide the natural choice as quantum information carrier for the client and enable quantum computing for the server. This is a unique feature of photonic systems and so far not realizable in other quantum systems.
We experimentally demonstrate the concept of BQC
via a series of blind computations on four-qubit blind cluster states.
As shown in Figure~\ref{figure1}, these photonic states can be combined via optical gates to create
a \emph{universal} resource state for BQC~\cite{Broadbent2009}. 

Our protocol uses, compared to the original BQC proposal~\cite{Broadbent2009}, the experimental resources in an optimised way, independent of the physical system and without affecting blindness.
%
\begin{figure*}
\includegraphics[width=0.90\textwidth]{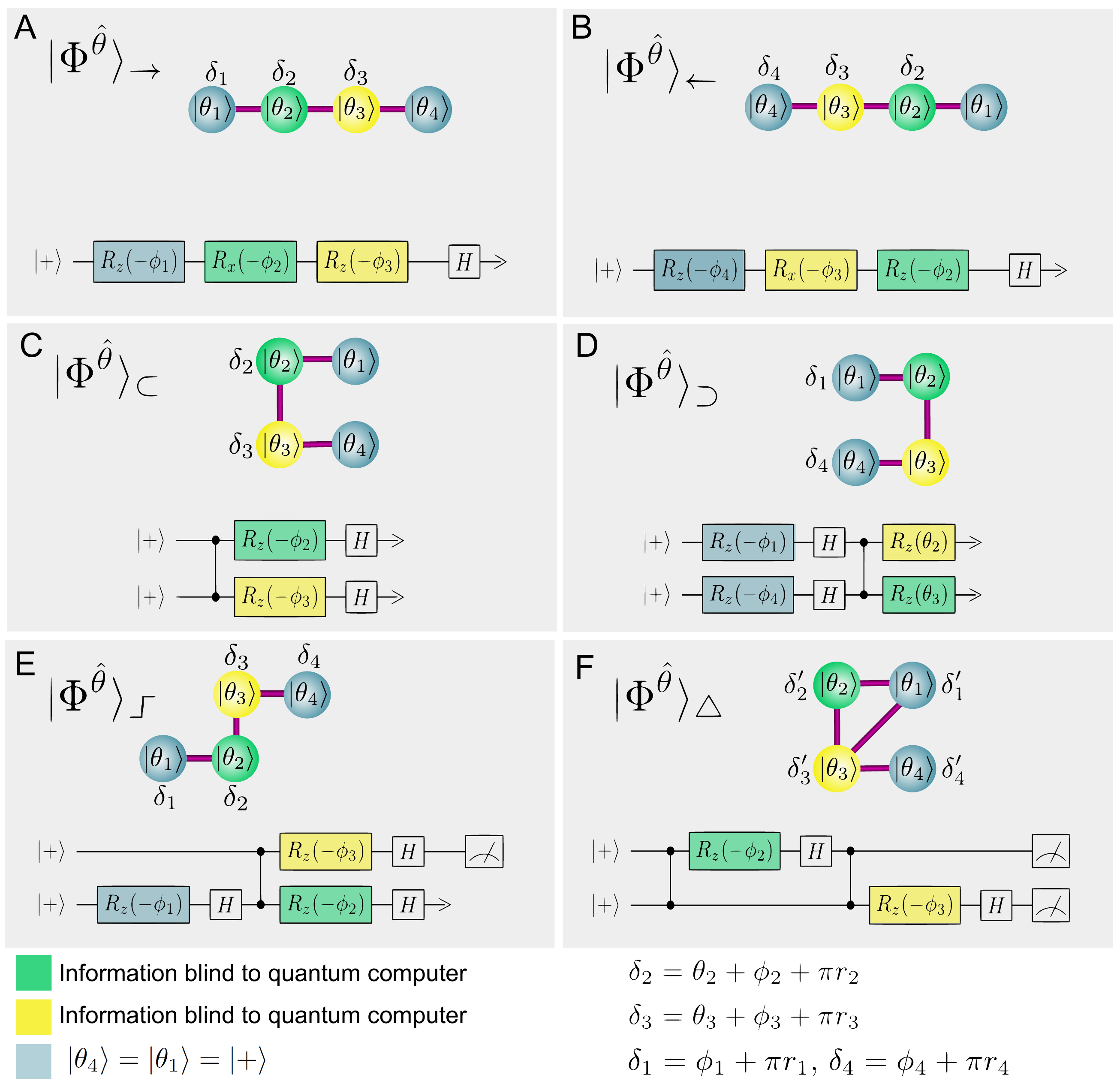}
\caption{\label{figure2}Blind circuits and corresponding measurement patterns. 
(\textbf{A-F}) We implement various types of blind computations using different configurations for $\ket{\Phi^{\hat{\theta}}}$. For all implementations, $\theta_2$ and $\theta_3$ are blind, as has been demonstrated in the experiment. The angles $\theta_1$ and $\theta_4$ are fixed to be zero.
Dependent on the experimental setting, we can implement blind linear cluster states $\ket{\Phi^{\hat{\theta}}}_{\rightarrow}$ (A) and $\ket{\Phi^{\hat{\theta}}}_{\leftarrow}$ (B), blind horseshoe $\ket{\Phi^{\hat{\theta}}}_{\subset}$ (C) and rotated-horseshoe cluster states $\ket{\Phi^{\hat{\theta}}}_{\supset}$ (D), blind staircase cluster states $\ket{\phi^{\hat{\theta}}}_{\stair}$(E) and blind triangle cluster states $\ket{\Phi^{\hat{\theta}}}_{\triangle}$(F).
The measurement angle $\delta_j$, as instructed by the client, depends on the initial rotation of the qubit $\theta_j$ (unknown to the server), the target rotation $\phi_j$ and a randomly chosen value $r_j$ in $\left\{0,1\right\}$.}
\end{figure*}
\section*{Optimised blind quantum computing}

It is a conceptual strength of the BQC protocol that perfect security can be established over a subset of computations even if not all of the qubits are unknown to the server. In fact, for the four-qubit blind cluster state it is sufficient for the client to be able to prepare only one or two of the qubits in arbitrary states $\ket{\theta_j}$  for delegating various one- and two-qubit circuits as well as quantum algorithms, see Figure~\ref{figure2}. This is a remarkable optimisation for the experimental requirements and is demonstrated for the first time here (see Appendix for theoretical details). Furthermore this optimisation is scalable beyond our four-qubit experimental setting
and creates an interesting challenge on the design level to construct a computation such that the sensitive measurements remain hidden.

We thus fix~$\theta_1$ and $\theta_4$ equal to zero, while varying the
choices of $\theta_2$ and $\theta_3$. 
The resulting four-qubit linear blind cluster state is:
\begin{eqnarray}
\label{eq:cluster}
  \ket{\Phi^{\hat{\theta}}}&=&\frac{1}{2}(\ket{+00+}_{1234} + e^{i\theta_3}\ket{+01-}_{1234}\\ \nonumber  
    &+& e^{i\theta_2}\ket{-10+}_{1234} - e^{i(\theta_2 + \theta_3)}\ket{-11-}_{1234})\,.
\end{eqnarray}

Our experimental implementation of BQC is based on such a family of four-qubit linear blind cluster states. These are produced using photon emissions of a non-collinear type-II spontaneous parametric down-conversion process (SPDC)~\cite{Kwiat1995, Walther2005a}, as described in detail in the Appendix.
If four photons are emitted into the output modes of the polarizing beam splitters 1, 2, 3 and~4 (see Figure 3A), they are in a highly entangled state which
is equivalent to the state $\ket{\Phi^{\hat{\theta}}}$ under the local unitary operation $H \otimes \mathbf{I} \otimes \mathbf{I} \otimes H$:
\begin{eqnarray}
\label{eq:lab-cluster}
  \ket{\Phi^{\hat{\theta}}_L}&=&   \frac{1}{2}(\ket{0000}_{1234} + e^{i\theta_3}\ket{0011}_{1234}   \\ \nonumber  
  &+&e^{i\theta_2}\ket{1100}_{1234} - e^{i(\theta_2 + \theta_3)}\ket{1111}_{1234})
\end{eqnarray}
where $\hat{\theta} = (n_2,n_3)$ and $(\theta_2, \theta_3)=(\frac{n_2\pi}4 , \frac{n_3\pi}4)$. In the experiment, we use the polarization of photons to represent the qubits, with $\ket{0}$ denoting the horizontal polarization state and $\ket{1}$ denoting the vertical polarization state.

The client prepares the value of $\theta_j$, which is done in our case by a human client.
By aligning our setup to produce $\ket{\Phi^{(2, n)}_L}$ for $n =
0,\ldots,7$ and $\ket{\Phi^{(6, m)}_L}$ for $m=0,4$, we have
demonstrated for the first time the 
preparation of various four-qubit blind cluster states. 
Moreover, we have implemented $1962$ different four-qubit measurements with 31392 measured outcomes.
These measurements outcomes can be seen as
implementing all possible computational branches (due to different
measurement outcomes), which is equivalent to directly performing the
feed-forward mechanism~\cite{Prevedel2007a}. 
However, the remarkably feature of the BQC protocol is that the client's privacy is always preserved, whether or not feed-forward mechanisms have been implemented.
Similarly, obtaining all the possible measurement outcomes is equivalent to implementing all possible values of $r_j$, as if the
client randomly re-interprets the measurement outcomes, implicitly
subsuming~$r_j$. Note
that, whenever all qubits are measured in our setup, this method
allows the client's choice of configuration to also be hidden from the
server.  

We use an over-complete state tomography for each of our cluster states, in order to reconstruct the four-qubit density matrix.
The most likely physical density matrix for each four-qubit state is extracted using a maximum-likelihood reconstruction~\cite{James2001} (see Figure~\ref{figure3}B).  Uncertainties in quantities extracted from these density matrices are calculated using a Monte Carlo routine and assumed Poissonian errors. Our computed fidelities for the various blind cluster states achieve maximum values of up to $67.9 \pm 0.4\%$ via local unitary transformation.
These non-ideal fidelities arise due to experimental imperfections (see Appendix).
It is important to note that experimental influences on the server's side only affect the correctness of the computation, while imperfections in the client's qubit preparation might also weaken the assumption of an unbiased state distribution.

\section*{Blind single- and two-qubit unitaries}
\label{sec:single-qubit gate}
\begin{figure*}
\includegraphics[width=0.85\textwidth]{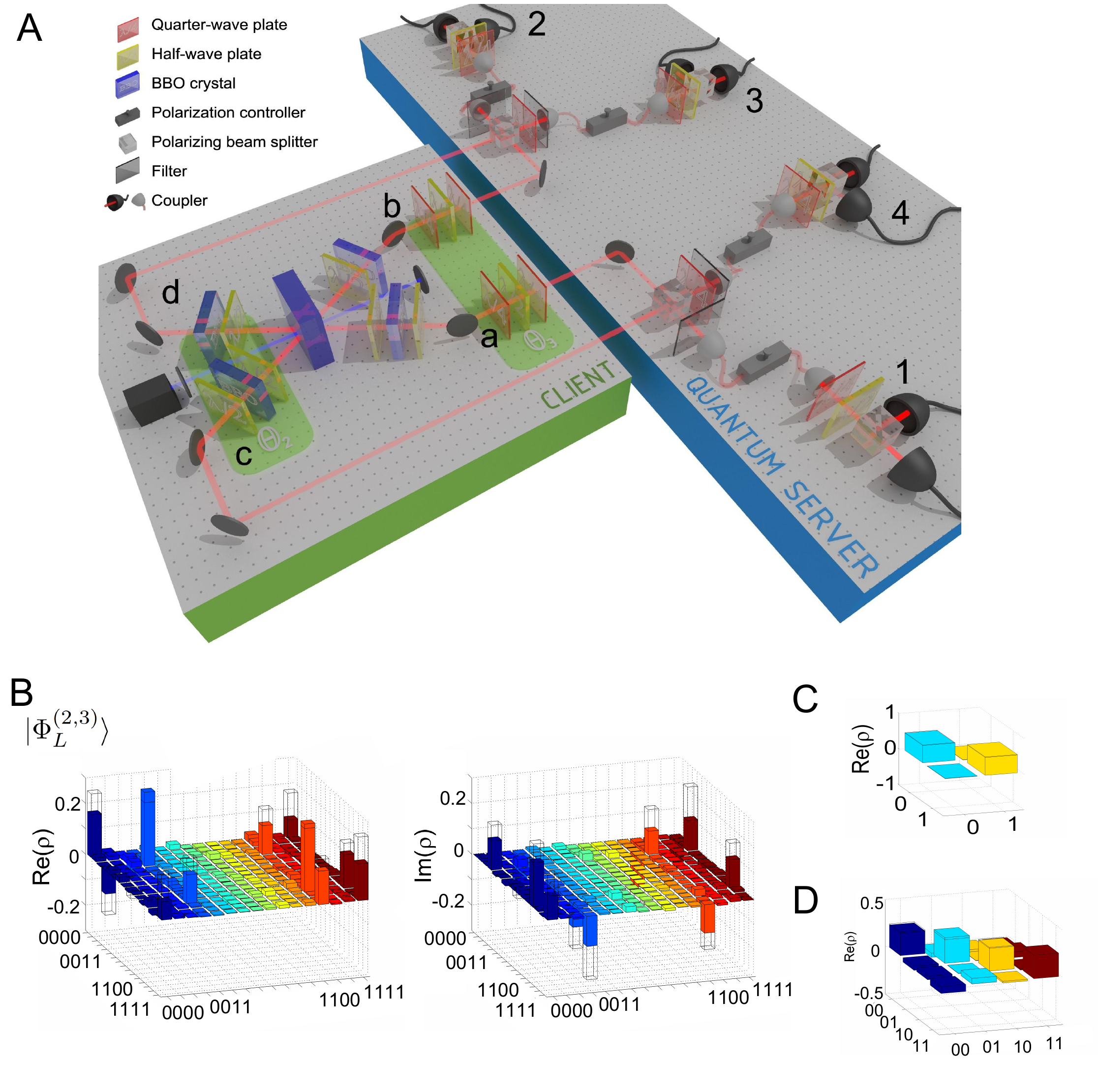}
\caption{\label{figure3}Experimental setup, measurement results (solid) and ideal values (wireframe). (\textbf{A}) The experimental setup to produce (client) and measure (quantum server) blind cluster states. The various blind cluster states are created by adjusting the settings of the phase retarders located along the path of the state emitted into the forward ($\theta_3$) and backward ($\theta_2$) modes (see Appendix for detailed information).
(\textbf{B}) Density matrix of the four-qubit cluster state $\ket{\Phi^{(2,3)}_L}$ in the laboratory basis. Shown are the real (left) and imaginary (right) parts of the density matrix. 
(\textbf{C}) Experimental demonstration of a single-qubit rotation around the Z-axis of the Bloch sphere and its consistency with blindness. 
A measurement of $\delta_4=\pi/2$, $\delta_3=-\pi/2$, and $\delta_2=-\pi/2$ on the blind linear cluster $\ket{\Phi^{(2,n_3)}}_{\leftarrow}$ results in rotations $\ket{\Psi_\text{out}}= R_x(\pi)\,R_z(\theta_3+\pi/2)\ket{\Psi_\text{in}}$ on the encoded qubit that depend on the initial rotation $\theta_3$.
By varying $\theta_3$ and averaging over all resulting density matrices, we obtain a totally mixed state. 
(\textbf{D}) Experimental demonstration a two-qubit gate and its consistency with blindness. 
Measuring $\delta_2=0$, and $\delta_3=-\pi/2$ at the states $\ket{\Phi^{(2,0)}}$, $\ket{\Phi^{(2,4)}}$, $\ket{\Phi^{(6,0)}}$, $\ket{\Phi^{(6,4)}}$ results in computations $R_z(\theta_2)\otimes R_z(\theta_3+ \pi/2)\mbox{ Cphase}\ket{\Psi_\text{in}}$ dependent on $\theta_2$ and $\theta_3$.
Averaging over this subset of all 64 possible states results in a totally mixed state. 
The imaginary part of the density matrices (C,D) is below 0.05 and hence not shown.}
\end{figure*}

The four-qubit linear blind cluster $\ket{\Phi^{\widehat{\theta}}}_\rightarrow$(Figure~\ref{figure2}) can be used to implement an arbitrary single-qubit unitary gate. Measuring qubit~1 in the eigenstates of $\sigma_x$, $\sigma_y$, or $\sigma_z$ has the effect of preparing the input on qubit~2 in the state $\ket{0}$, $\ket{+_i}$ or $\ket{+}$, respectively, where $\ket{+_i}=1/\sqrt{2}(\ket{0} + i\ket{1})$.
We are thus left with a three-qubit linear cluster state that implements a single-qubit rotation gate with rotations determined by the measurements of the second and third qubits; this rotates the input qubit $\ket{\Psi_{\text{in}}}$ to the final state $\ket{\Psi_{\text{out}}}= R_x(-\phi_3)R_z(-\phi_2)\ket{\Psi_{\text{in}}}$, where $R_x(\alpha) = \exp(-i\alpha \sigma_x /2)$. 
By fixing $\theta_2$ and varying $\theta_3$, 
we can experimentally demonstrate a blind $X$-rotation. 
In the same way, a blind $Z$-rotation can be shown experimentally by using the four-qubit linear blind cluster state $\ket{\Phi^{\widehat{\theta}}}_{\leftarrow}$, which has the order of measurements going from qubit~4 down to qubit~1. 
Figure~\ref{figure3}C depicts an experimental demonstration of a blind $Z$-rotation. 
By varying $\theta_3$ and averaging over all resulting density matrices, we obtain a totally mixed state with a linear entropy of $0.989\pm0.010$ that is close to the entropy of 1 for a perfectly mixed state (see Figure~\ref{figure3}C). 
As the experiments include the preparation of
all eight blind cluster states $\ket{\Phi_L^{(2,n)}}$, we can quantify
the blindness of the single-qubit rotations demonstrated experimentally. The value of the Holevo information $\chi$ (see Appendix for details) must then be between $0$ (for perfect blindness) and $3$ (for no blindness). Using the tomographic measurements
performed on these input states we determine $\chi$
of such states to be $0.169\pm 0.074$, far below the three bits
necessary to uniquely identify the client's choice of $\phi_{2}$ and $\phi_{3}$,
proving that within the assumptions of our model these experimental
implementations of the protocol maintains close to perfect
blindness.
The above value of $\chi$ assumes initial state is chosen uniformly at random. However even when this value is maximised over all possible prior distributions on the choice of states, it increases only slightly to $0.185 \pm 0.087$.

Two-qubit gates are required for universal quantum computation; by choosing the order of measurements in a suitable way, the blind cluster $\ket{\Phi^{\hat{\theta}}}$ implements blind two-qubit gates (Figure~\ref{figure2}C--F).
One family of two-qubit gates generated in our experiment is based on the blind horseshoe cluster $\ket{\Phi^{\hat{\theta}}}_{\subset}$, where measuring qubits 2 and 3 of the blind cluster state performs a transformation on the logical input qubits (Figure~\ref{figure2}C). Both implemented rotations are blind and the entire computation remains hidden. Analyzing the output state, \emph{i.e.}~measuring qubits 1 and 4, delivers the result of the computation. 
Figure~\ref{figure3}D shows an example of a two-qubit computation using the blind horseshoe cluster.
Consistency with blindness can be seen by averaging over all output states, giving as a result a totally mixed state with a linear entropy of $0.955\pm0.011$. 
It is an interesting challenge to demonstrate the consistency with blindness in full generality by producing 64 blind cluster states. Our demonstration uses a selection of four states which suffices to hide the choice of rotations among four possibilities:  $R_z(\pi/2 \pm \pi)\otimes R_z(\pi/2 \pm \pi)$. 
In a similar way, the consistency with blindness of the rotated horseshoe cluster $\ket{\Phi^{\hat{\theta}}}_{\supset}$ (Figure \ref{figure2}D) can be shown. 
We also realize blind computations based on the blind staircase cluster $\ket{\Phi^{\hat{\theta}}}_{\stair}$  (Figure~\ref{figure2}E) and blind triangle cluster~$\ket{\Phi^{\hat{\theta}}}_{\triangle}$ (Figure~\ref{figure2}F). 
The state~$\ket{\Phi^{\hat{\theta}}}_{\triangle}$ is obtained via local complementation~\cite{VandenNest2004} on qubit~2 of $\ket{\Phi^{\hat{\theta}}}_{\subset}$. Thus $U\ket{\Phi^{\hat{\theta}}}_{\triangle}=\ket{\Phi^{\hat{\theta}}}_{\subset}$ where $U=\sqrt{\sigma_z}\otimes\sqrt{\sigma_x}\otimes\sqrt{\sigma_z}\otimes I$ (with $U$ acting on qubits ordered as 1,2,3,4), and
measuring the qubits of $\ket{\Phi^{\hat{\theta}}}_{\subset}$ by absorbing the action of $U$ into the measurements  yields a computation on  $\ket{\Phi^{\hat{\theta}}}_{\triangle}$ as represented by measurement instructions $\delta'$ in Figure~\ref{figure2}F. However, blindness on qubit~$i$ is guaranteed only if the resulting measurement can be expressed as a basis~$\ket{\pm_{\delta_i'}}$.
We show in the next sections the blind staircase cluster allows for the blind implementation of Deutsch's algorithm, while the blind triangle cluster allows for the blind implementation of Grover's algorithm.

\section*{Blind algorithms}
\label{sec:Grover}
\begin{figure}
\includegraphics[width=0.9\columnwidth]{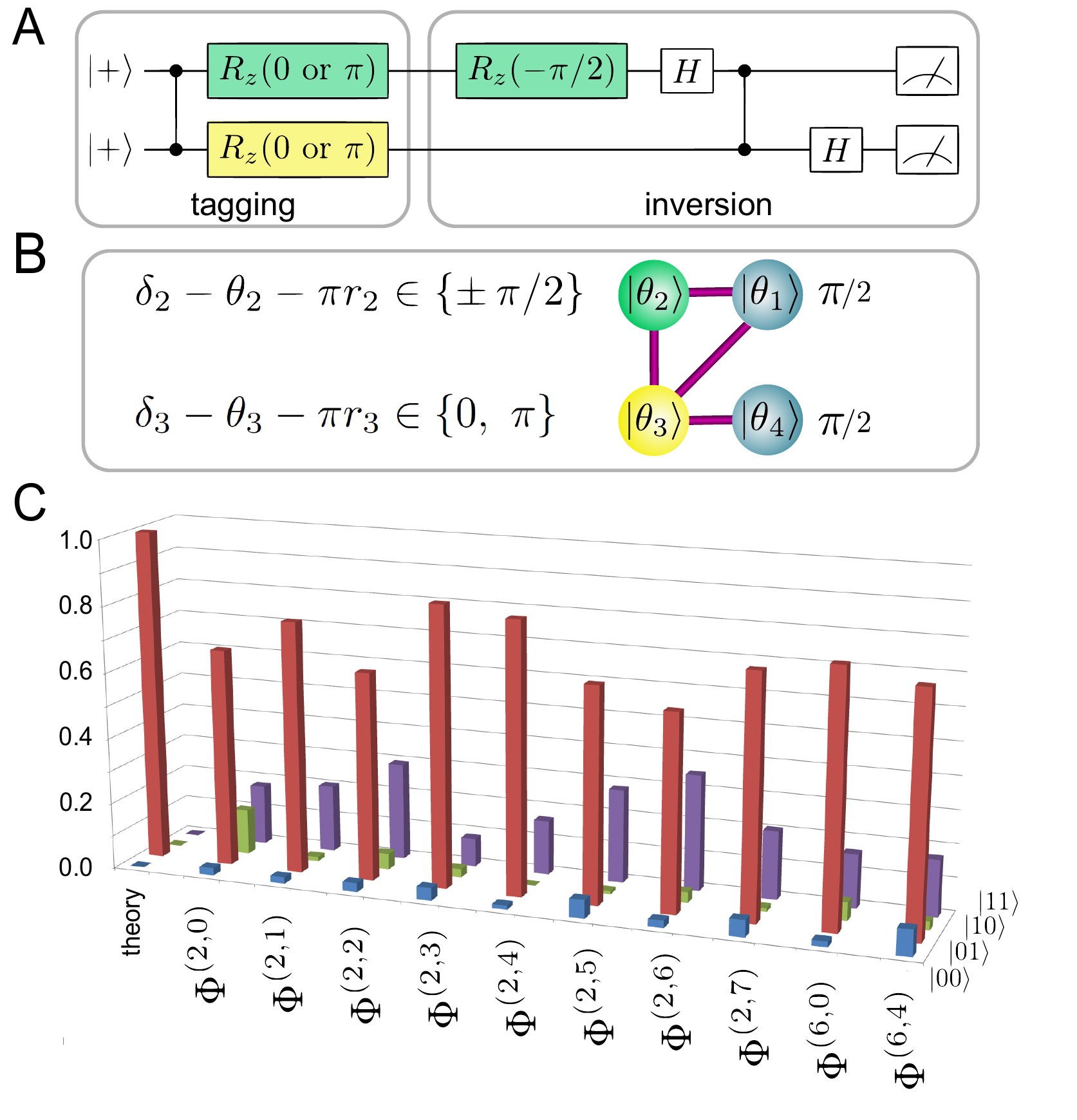}
\vspace{-0.5cm}
\caption{\label{figure4}Blind implementation of Grover's algorithm. (\textbf{A}) Quantum circuit. The input to the circuit is $\ket{+}\ket{+}$; applying one of the operations $R_z(0\,\,\text{or}\,\, \pi)\otimes R_z(0\,\,\text{or}\,\, \pi)\mbox{CPhase}$  defines which of the four input states $\ket{00}$,$\ket{01}$,$\ket{10}$,$\ket{11}$ is tagged and applies a phase shift of $\pi$ to that state. The operation $(I\otimes H) \mbox{CPhase}(H R_z(-\pi/2)\otimes I)$ then maps these four states to an output that is measured in the basis ($\ket{+_i},\ket{-_i}$). (\textbf{B}) Corresponding implementation on a triangle cluster $\ket{\Phi^{\hat{\theta}}}_{\triangle}$. 
Here, the measurement of qubits 2 and 3 corresponds to the tagging of one of the elements, measuring the output qubits 1 and 4 with measurement angles of $\pi/2$ identifies then which input was tagged. Depending on the state we want to tag, we choose one set of measurement angles on qubits 2 and 3 from the four possible sets given in Figure~\ref{figure4}a. For example,  a measurement  with angles $-\pi/2$ and $\pi$ tags the state $\ket{01}$. 
Since qubits 2 and 3 are blind, the measurement instructions depend on the initial rotation of the qubit. Without that knowledge, the quantum server is unable to distinguish the algorithm from a given family of circuits. (\textbf{C}) Measurement outcomes for tagging the $\ket{01}$ element for all states $\ket{\Phi^{(n_2,n_3)}}_{\triangle}$ are shown. The corresponding error bars are smaller than $0.056$ for all results shown.}
\vspace{-0.1cm}
\end{figure}

One of the most prominent examples where quantum mechanics
demonstrates its superiority in computational speedup
is Grover's search algorithm~\cite{Grover1996, Boyer1998}, which provides a quadratic speedup to the following problem:
Given a function~$f :\{0,1\}^n \rightarrow \{0,1\}$, find an $x$ such that $f(x)=1$.  
Here we demonstrate a blind implementation of Grover's search for $n=2$, where blindness ensures that the server is unable to distinguish the actual computation from within a given family of circuits implementing~$(I\otimes R_z(\xi)H)$. 
Whereas previous realisations~\cite{Walther2005a, Prevedel2007a} are not amenable to blind implementations, our computation, embedded into the blind triangle cluster $\ket{\Phi^{\hat{\theta}}}_{\triangle}$ (Figure~\ref{figure2}F), 
remains blind.
The algorithm  proceeds as follows: the values of~$x$ are represented by  the states $\ket{00}$, $\ket{01}$, $\ket{10}$ and $\ket{11}$, respectively. A superposition of all four states is initially created and the oracle tags one element by  applying a phase of~$\pi$, thus flipping the sign of this term (see Figure~\ref{figure4}A). Then each
of the four states is mapped to an output such that measuring both qubits in the basis $\ket{\pm_i}$ reveals the tagged item. This computation can be embedded into the blind triangle cluster, $\ket{\Phi^{\hat{\theta}}}_{\triangle}$ (Figure~\ref{figure2}F), the choice of $\phi_2$ and $\phi_3$ determines which element is tagged.
Figure~\ref{figure4}C shows the results of a Grover search for the tagging of the state $\ket{01}$. For each blind cluster state, we show the probability of identifying the tagged state as well as the probabilities of finding the unwanted states, due to the experimental noise.
We achieve probabilities of finding these positive events of up to $0.850\pm0.039$ with an average over all blind states of $0.720\pm0.015$. Note that no classical algorithm can succeed in this scenario with probability higher than~$0.5$.

\begin{figure}
\includegraphics[width=0.95\columnwidth]{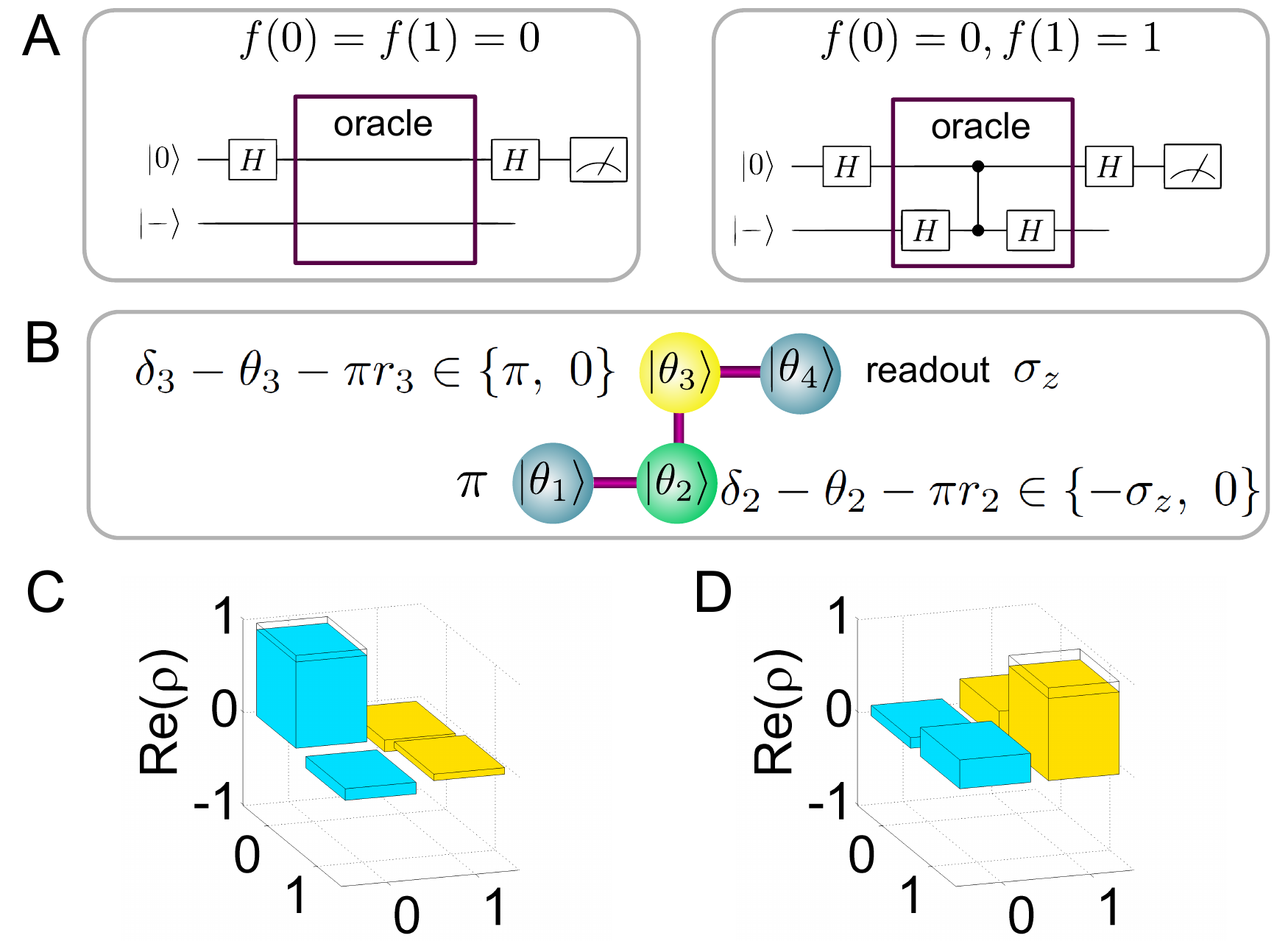}
\caption{\label{figure5} Blind implementation of Deutsch's algorithm. (\textbf{A},\textbf{B}) The quantum circuits
and the corresponding measurements on a staircase cluster state
$\ket{\Phi^{\hat{\theta}}}_{\stair}$ for the constant and the balanced
oracle, distinguished by the measurement of qubit 2 and qubit 3. Blindness of qubit 3 guarantees that the quantum server cannot distinguish between the execution of each of these scenarios (constant or balanced oracles) and corresponding families of quantum circuits.
(\textbf{C}, \textbf{D}) Experimental (solid) and theoretical (wireframe) results for a constant (C) and a balanced (D)
oracle for the example of the state $\ket{\Phi^{{(6,4)}}}_{\stair}$.}
\vspace{-0.1cm}
\end{figure}

Another algorithm which demonstrates the power of quantum computing, is the Deutsch-Josza algorithm~\cite{Deutsch1992} that takes as input an oracle (or black-box) for computing an unknown function $f: \{0,1\}^n \rightarrow \{0,1\}$ with the promise that $f$ is either \emph{constant}, meaning $f(x)$ is the same for all~$x$, or \emph{balanced}, meaning $f(x)=0$ for exactly half of the inputs $x$ and $f(x)=1$ for the other half. The algorithm determines whether $f$ is constant or balanced by making queries to the oracle. While the best possible classical algorithm to solve this problem  uses at least $2^{n-1} + 1$ queries in the worst case, the Deutsch-Jozsa algorithm takes advantage of quantum superposition and interference to determine whether $f$ is constant or balanced with only \emph{one} query. 
In contrast to previously realized implementations of Deutsch's algorithm using traditional cluster states~\cite{Tame2007, Vallone2010}, we exploit blind staircase cluster states $\ket{\Phi^{\hat{\theta}}}_{\stair}$ for the implementation of this quantum algorithm for the case $n=1$. 
Figure~\ref{figure5}A shows the quantum circuits that realise oracles corresponding to constant and balanced functions.
The corresponding implementation on $\ket{\Phi^{\hat{\theta}}}_{\stair}$ is given in Figure~\ref{figure5}B, where the choice of oracle is done by fixing the measurement on qubits~2 and~3. 
Blindness of qubit 3 guarantees that the quantum server will not recognize the implementation of a constant oracle from the Grover algorithm or general circuits implementing ~$R_z(\xi)H \otimes I$, and a balanced oracle from $(I\otimes H) \mbox{CPhase}(R_z(\xi)H\otimes H)$, see Figure~\ref{figure5}. 

Figure~\ref{figure5}C and~\ref{figure5}D show the outcome of our measurement for the case of $\ket{\Phi^{(6,4)}}_{\stair}$. A tomography of the state of qubit~4 is performed in order to fully characterize the output of the computation. In this case, the obtained fidelity for the output state is $F=0.930 \pm 0.025$ for the constant oracle and $F=0.887 \pm 0.033$ for the balanced oracle, with the algorithm producing the correct result with probabilities $0.899 \pm 0.006$ for the constant and $0.895 \pm 0.022$ for the balanced oracle.

\section*{Towards verifying the quantumness}
\label{sec:testing}
Self-testing is a verification process for the operations of a collection of untrusted quantum devices~\cite{Magniez2006, McKague2011}; a key application of the blind computing protocol is also towards such verification of quantum devices~\cite{Broadbent2009,Aharonov2010}. In the setting of our experiment we demonstrate a notion of verification that can be used as a heuristic probabilistic test for whether the server indeed possesses any quantum technology or is a completely classical device.
For this, the client chooses a measurement setting for which, for each measurement outcome, there exists a state with a detection probability of zero.
Due to blindness, however, the quantum server has no information about which initial states the client has prepared.
If it has no quantum technology in hand, it attempts to use its classical devices and guesses the outcome for the client's computation wrong with probability at least 1/8.
Better bounds can be achieved using statistics of several rounds by comparing it with the known theoretical statistics to test
whether the quantum-computing server is producing the expected outcome or not.
%
\begin{figure}
\includegraphics[width=0.85\columnwidth]{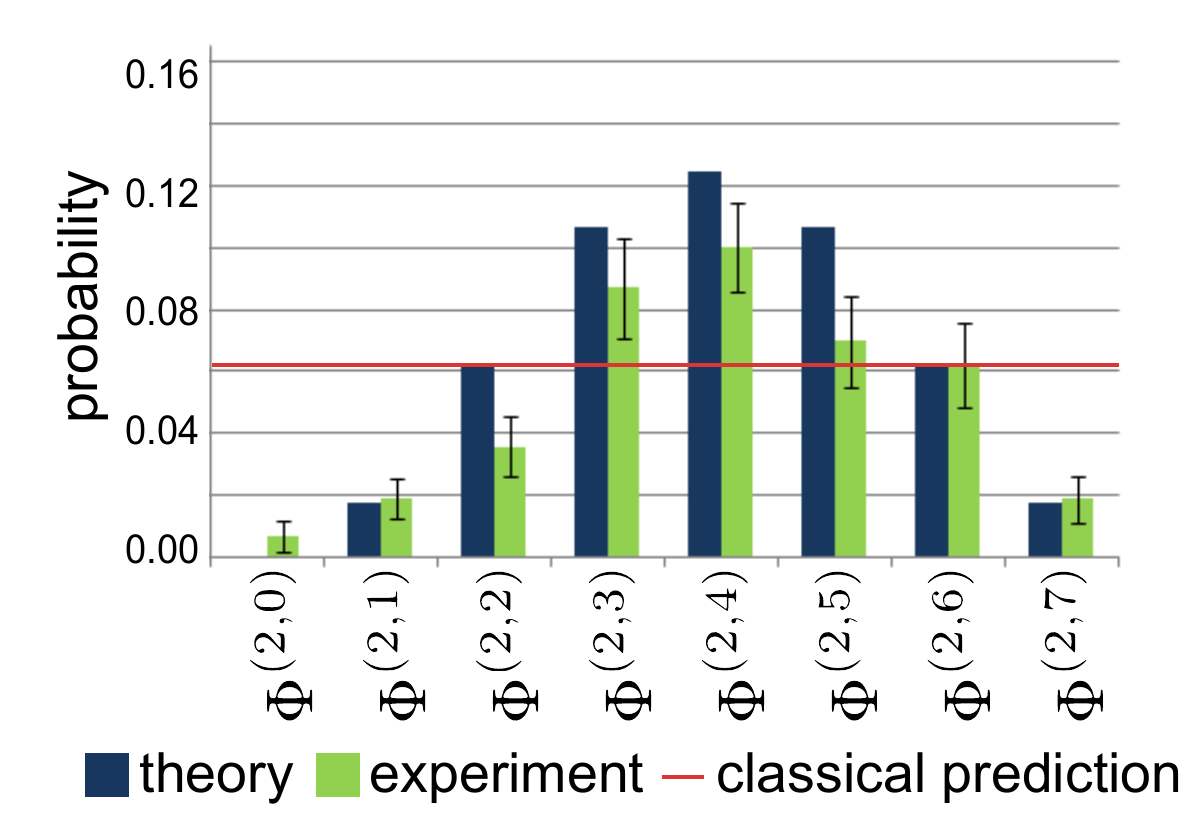}
\caption{\label{figure6}Testing of the quantum server. 
By measuring the probability distributions of a fixed measurement setting for all blind cluster states and comparing those with the theoretical expectations, the client can find out if the server possesses any quantum technology or not. For example, fixing the measurement settings to $\delta_1=-\sigma_z$, $\delta_2=\pi$, $\delta_3=-\pi/2$, and $\delta_4=\pi/2$ leads to different theoretical (blue) and experimental (green) probability distributions dependent on the underlying blind cluster state. A pure classical server guesses every outcome with the same probability ($1/16$) and can be detected in this way (red line).
Conservatively, we show Poissonian errors which constitute a lower limit for the experimental error due to imperfections in the state generation (see Appendix).
}
\end{figure}

We experimentally demonstrate the testing procedure using statistics of several outcomes for different measurement instructions. Figure~\ref{figure6} shows relevant theoretical predictions as well as experimental outcomes which confirm the quantum nature of the server. By instructing the quantum server to measure, for example, this statistical distribution, the client can see if the outcomes coincide with the expectations. Our demonstration is the first step towards an efficient verification scheme for quantum technology and acts as experimental benchmark for future fault-tolerant protocols using more qubits, that is expected to enable the detection of a cheating quantum server with probability exponentially close to one.

\section*{Discussion}
\label{sec:discussion}

We have experimentally demonstrated the concept of blind quantum computing. Generating a family of four-qubit blind cluster states, we obtained a universal set of single-qubit and non-trivial two-qubit quantum logic gates, as well as implementations of Deutsch's and Grover's algorithms. We use the photon's mobility, an intrinsic advantage of this physical quantum system, for preparing various qubits on the client's side which are then processed by a locally separated quantum server. 

On the path from our proof-of-principle experiment to a full implementation of the BQC scheme, there are several technical challenges to be faced: Emitted photons that do not contribute to the generation of the cluster state can in principle reveal information about the blind phases. Furthermore, post-selection and photon losses decrease the efficiency of the protocol. Therefore, the realization of single-qubit states on demand and the heralded generation of blind cluster states using measurement-induced interactions with high fidelity and low losses will be crucial for future applications. In our experiment, the blind angles were chosen by the human client and the measurement settings were selected from a prepared list. Ideally, the source of randomness should be carefully scrutinized to avoid any correlations with the server, and an efficient shot-by-shot randomization should be implemented. Considering the photon rates in our experiment, the realization of full randomization for each measurement is a major challenge, but seems within reach by advancing current technologies. The question of how far imbalances and deviations from the uniform distribution can be acceptable is a topic of current research.

Our experiment is the first step towards unconditionally secure quantum computing in a client-server environment, where the client's entire computation remains hidden --- a functionality not known to be achievable in the classical world. We anticipate that this will become an important privacy-preserving technique in future quantum computing networks or clouds~\cite{Hayes2008}. Especially considering the tremendous challenges encountered in making quantum computers widely available, such future networks could consist of a few powerful quantum-computer nodes. The only quantum requirement for the clients would be to communicate with the nodes via quantum links enabling the transfer of arbitrary qubits. Although photonic quantum systems seem to be ideally suited for privacy-preserving quantum computing, we stress that our results are applicable to any physical implementation of qubits and that in the near future the precise quantum control of multi-qubit quantum systems~\cite{Monz2011} will allow for implementing more complex algorithms.

The authors are grateful to C.~Brukner, V.~Danos and R.~Prevedel for discussions, and to F.~Cipcigan and J.~Schm\"{o}le for support.
We acknowledge support from the European Commission, Q-ESSENCE (No~248095), 
ERC senior grant (QIT4QAD), JTF, Austrian Science Fund (FWF): [SFB-FOCUS] and [Y585-N20], EPSRC, grant EP/E059600/1,
Canada's NSERC, the Institute for Quantum Computing, QuantumWorks, the National Research Foundation and Ministry of Education, Singapore,
and from the Air Force Office of Scientific Research, Air Force Material Command, USAF, under grant number FA8655-11-1-3004.



\newcommand{\SortNoop}[1]{}


\section{Appendix}

\subsection{Proof of blindness for optimised BQC}
Our optimised protocol guarantees full blindness of a computation, even though not all qubits have an a initial rotation.
Here, we show that algorithms that admit a measurement
pattern in which the secret can be encoded over a few qubits can be
made fully blind using only a few blind qubits. This construction
works as long as measurements following the blind qubit measurements
are chosen in the Clifford group (integer multiples of $\pi/2$), and 
thus our constructions works for Deutsch's and Grover's algorithms.

In our setup, in order to achieve blindness for measurements of
qubits 2 and 3 the quantum server should learn nothing about
$\phi_2$ and $\phi_3$, the client's choice of measurements on those
qubits. The quantum server holds the rotated qubits
$\frac{1}{\sqrt{2}}\left(\ket{0}+e^{i\theta_2}\ket{1}\right)$ and
$\frac{1}{\sqrt{2}}\left(\ket{0}+e^{i\theta_3}\ket{1}\right)$; it
also has the measurement instructions $\delta_2 = \phi_2+\theta_2 +
\pi r_2$ and $\delta_3 = \phi_3+\theta_3 + \pi r_3$. For simplicity
we define new variables $\tilde{\theta}_2 = \theta_2 + \pi r_2$ and
$\tilde{\theta}_3 = \theta_3 + \pi r_3$. Thus the server holds
the quantum states
$\frac{1}{\sqrt{2}}\left(\ket{0}+e^{i(\tilde{\theta_2} + \pi
r_2)}\ket{1}\right)$ and
$\frac{1}{\sqrt{2}}\left(\ket{0}+e^{i(\tilde{\theta}_3 + \pi
r_3)}\ket{1}\right)$. Since $r_2$ and $r_3$ are random and unknown to
the quantum server, the density matrices corresponding to these
systems, as held by the server, are identically $\mathbb{I}/2$, that
is, the systems are completely mixed and thus independent of
$\tilde{\theta}_2$ and $\tilde{\theta}_3$. As these angles are
themselves uniformly random, the classical information $\delta_2 =
\phi_2 + \tilde{\theta_2}$ and $\delta_3 = \phi_3 +
\tilde{\theta_3}$ is also uniformly random and independent of
$\phi_2$ and $\phi_3$. Hence, the quantum server, despite receiving
classical information $\delta_2$ and $\delta_3$ and quantum states
$\ket{\theta_2}$ and $\ket{\theta_2}$, cannot distinguish between
the possible choices of the client's measurements on qubits 2 and 3.
Now, these blind measurements may be followed by non-blind ones.
This will not affect the blindness (no information about $\phi_2$
and $\phi_3$ will be leaked to the server) as long as the structure
of the algorithm permits the non-blind angles to be integer
multiples of $\pi/2$.
This optimisation is due to the fact that for $\phi_4 \in \{0, \pi/2, \pi, 3\pi/2\}$, a sign flip on $\phi$ ($\phi\ \rightarrow\ -\phi$) can be re-interpreted as the (possible) addition of $\pi$ to $\phi\ (\phi\ \rightarrow\ \phi\ +\ \pi)$. Thus the feedforward structure for $\phi_4$ is given by the addition of an optional multiple of $\pi$. This is completely hidden by the usual addition of a random $\pi r_4$, included in $\delta_4$. Thus, from the server's point of view, the process of measuring qubit~4 is independent of $\phi_2$ and~$\phi_3$.  

\subsection{Leakage of information in experimental blind quantum computing}
In order to quantify any deviation from perfect blindness
introduced by experimental imperfections we need a model for the
information received by the server during a run of the protocol. To
this end, we assume that the only information obtained by the server
is the initial quantum state supplied by the client, as well as the
classical set of angles $\delta_i$ received during the protocol. We
also assume that the state produced by the experimental setup for a
given choice of the ideal input state is fixed, and that the
client's choices can be considered uniformly random. In such a
model, the amount of information leaked is bounded by the Holevo
information of the quantum state, $\rho$ received by the server,
$\chi = -\mbox{Tr}(\rho \log_2 \rho) + \sum_{\theta} \frac{1}{8}
\mbox{Tr}(\rho_\theta \log_2 \rho_\theta)$, where $\rho_\theta =
\frac{1}{2}(\rho'_\theta +\rho'_{\theta + \pi})$ and $\rho'_\theta$
is the state produced in the experimental apparatus for the client's
choice of $\theta$.

\subsection{Experimental setup}
In our experiment (main paper, Fig. 3A), entangled photon pairs are produced by exploiting the emissions of a non-collinear type-II SPDC process. For this, a mode-locked Mira HP Ti:Sa oscillator is pumped by a  Coherent Inc.~Verdi
V-10 laser. The pulsed-laser output ($\tau$ = $200\,$fs, $\lambda$= $789\,$nm, $76\,$MHz) is frequency-doubled using a $2\,$mm-thick Lithium triborate (LBO) crystal, resulting in UV pulses of $0.8\,$W cw average.
We achieve a stable source of UV-pulses by translating the LBO to avoid optical damage to the anti-reflection coating of the crystal.
Afterwards, dichroic mirrors are used to separate the up-converted light from the infrared laser light. The UV beam is focused on a $2\,$mm-thick $\beta$-barium borate (BBO) crystal cut for non-collinear type-II parametric down-conversion. Afterwards the beam is reflected and passes the crystal a second time. Entangled photon pairs are emitted into the forward modes, $a$ and $b$, and the backward modes, $c$ and $d$. Half-wave plates (HWPs) and additional BBO crystals compensate walk-off effects and allow the production of any Bell state in the forward and backward mode.
The modes of the different pairs $a$, $c$ and $b$, $d$, respectively, are then coherently overlapped at polarizing beam splitters (PBSs) by equalizing the different path lengths.
Narrow-band interference filters ($\Delta\lambda$ = $3\,$ nm) are used to spatially and spectrally select the down-converted photons which are then coupled into single-mode fibers that guide them to the polarization analysis setup. There, different polarization measurements are performed using quarter-wave plates (QWPs), HWPs and polarizing beam splitters.

\subsection{Experimental preparation of blind cluster states}
The blind cluster state in the laboratory basis is composed of four terms that correspond to different emissions of four photons. 
Such four-photon emissions can experimentally be obtained either by an emission of two entangled pairs, one in the forward and one in the backward mode, or by double-pair emissions into the forward or the backward mode.
The production of our cluster state exploits coherent superpositions of these different four-pair contributions and utilizes the properties of the polarising beam splitters (PBSs) as well as post-selection to obtain the appropriate state.
In order to produce the desired state, we align our setup such that a $\ket{{\Phi}^-_{\theta_3}}=(\ket{HH}-e^{i\theta_3}\ket{VV})/\sqrt{2}$
state is emitted in the forward direction and a $\ket{{\Phi}^+_{\theta_2}}=(\ket{HH}+e^{i\theta_2}\ket{VV})/\sqrt{2}$ state in the backward direction, where $\ket{H}$ ($\ket{V}$) denotes the horizontal (vertical) polarization state. The emission of only one entangled pair in the forward direction and only one pair in the backward direction  results in two different four-photon terms:
$\ket{H}_1\ket{H}_2\ket{H}_3\ket{H}_4$ and $- e^{i(\theta_2+\theta_3)}\ket{V}_1\ket{V}_2\ket{V}_3\ket{V}_4$ due to the properties of the PBSs. The two-pair emissions also lead to fourfold coincidences, namely to a $-e^{i\theta_3}\ket{H}_1\ket{H}_2\ket{V}_3\ket{V}_4$ state and a $e^{i\theta_2}\ket{V}_1\ket{V}_2\ket{H}_3\ket{H}_4$ state for a double-pair emission in the forward and in the backward direction, respectively.
We shift the phase of the term
$-e^{i\theta_3}\ket{H}_1\ket{H}_2\ket{V}_3\ket{V}_4$ by $\pi$ to generate a sign shift.
For this, we use the method~\cite{Walther2005a} where a rotation of an additional wave plate has the desired effect. The final output state is a superposition of all these four terms.
In our setup we use a combination of quarter-wave plates and half-wave plates to adapt the phase of the state that is emitted into the forward mode. The phase of the backward pair is adapted by tilting one of the compensation crystals. Note that after the PBS two quarter-wave plates are inserted in modes 3 and~4 to compensate for birefringence effects.
By changing the phases of the entangled pairs, we can now manipulate the values of $\theta_2$ and $\theta_3$ of the client's qubits.

In our experiment, the emitted Bell pairs show a typical visibility of about $0.9$, depending on the specific experimental setting. The different photon emissions then interfere at the PBSs with average visibilities of $0.85$.  Additional errors arise due to phase drifts during the measurements.  These main error contributions, together with minor errors like polarisation drifts, decrease the fidelity of our blind cluster states with respect to the ideal state. In our calculations, we always assume Poissonian errors. In fact, these indicate a lower bound for the actual error that takes all the experimental imperfections into account. This is underlined by an analysis of the data of Figure~6 (main paper) where we obtain a value of $\chi^2$ of $1.6$ when assuming Poissonian errors only. Including the errors mentioned above, we obtain a $\chi^2$ value satisfactorily close to one (about $1.1$).

In the context of BQC, the client has access to the various four-photon emissions and prepares the encoded phases, for example $\theta_2$ and $\theta_3$, by applying local operations. These photons are then sent to the quantum server who generates entangled blind cluster states by superimposing these qubits on two polarizing beam splitters, followed by a successful detection of a four-fold coincidence in the output modes 1-4. The settings for the computation are set by phase retarders in each of the output modes to align the setting for the consecutive projective measurements.  We note that due to the down-conversion process the client rather prepares arbitrarily rotated Bell pairs instead of single qubits, which enables a compact client-server network without affecting the blindness.

\end{document}